\documentclass[prl,twocolumn, aps,amssymb,footinbib,showpacs]{revtex4-1}

\usepackage{graphicx}

\usepackage{amsmath}
\usepackage{braket}
\usepackage{graphicx} 
\usepackage{color}

\def\urlprefix{}
\def\url#1{}

\usepackage{array} 
\usepackage{paralist} 
\usepackage{verbatim} 
\usepackage{braket} 

\newcommand{\be}{\begin{equation}}
\newcommand{\ee}{\end{equation}}
\newcommand{\beq}{\begin{eqnarray}}
\newcommand{\eeq}{\end{eqnarray}}
\newcommand{\Heff}{H_{\mathrm{eff}}}
\newcommand{\nop}{\textbf{n}_{\theta_1,\theta_2}(\hat{k})}
\newcommand{\n}{\textbf{n}_{\theta_1,\theta_2}(k)}
\newcommand{\eps}{\epsilon_{\theta_1,\theta_2} (k)}
\newcommand{\epsop}{\epsilon_{\theta_1,\theta_2} (\hat{k})}

\begin{document}

\title{Direct Probe of Topological Invariants Using Bloch Oscillating Quantum Walks}

\author{V. V. Ramasesh$^1$}
\author{E. Flurin$^1$}
\author{M. Rudner$^2$}
\author{I. Siddiqi$^1$}
\author{N. Y. Yao$^1$}
\affiliation{$^1$Department of Physics, University of California, Berkeley CA 94720}
\affiliation{$^2$ Niels Bohr International Academy and Center for Quantum Devices,
University of Copenhagen, 2100 Copenhagen, Denmark}

\date{\today}

\begin{abstract}
The topology of a single-particle band structure plays a fundamental role in understanding a multitude of physical phenomena.
	Motivated by the connection between quantum walks and such topological band structures, we demonstrate that a simple time-dependent, Bloch-oscillating quantum walk enables the direct measurement of topological invariants. 
We consider two classes of one-dimensional quantum walks and connect the global phase imprinted on the walker with its refocusing behavior. 
By disentangling the dynamical and geometric contributions to this phase we describe a general strategy to measure the topological invariant in these quantum walks. 
As an example, we propose an experimental protocol in a circuit QED architecture where a superconducting transmon qubit plays the role of the coin, while the quantum walk takes place in the phase space of a cavity. 

\end{abstract}

\maketitle

Much like their classical stochastic counterparts, discrete-time quantum walks~\cite{Venegas-Andraca2012} have stimulated activity across a broad range of disciplines. 
In the context of computation, they provide exponential speedup for certain oracular problems and represent a universal platform for quantum information processing~\cite{PhysRevA.58.915,doi:10.1137/S0097539705447311,childs_universal_2009}. 
Quantum walks also exhibit features characteristic of a diverse set of physical phenomena, ranging from localization to molecule formation \cite{kitagawa_exploring_2010,preiss_strongly_2015}.  
At their core, discrete-time quantum walks (DTQW) are dynamical protocols associated with spinful particles on a lattice, where the  internal spin state controls the direction of motion \cite{nayak2000quantum,childs2004spatial,PhysRevB.92.045424,1751-8121-49-21-21LT01,kitagawa_exploring_2010,kitagawa2012observation,kitagawa2012topological,PhysRevB.88.121406,PhysRevB.86.195414,PhysRevA.89.042327,karski2009quantum,preiss_strongly_2015,zahringer2010realization,schreiber2010photons,schmitz2009quantum,kitagawa2012observation}. 
Motivated by this intrinsic spin-orbit coupling, a tremendous body of recent work has focused on exploring the  topological features of DTQWs both theoretically and experimentally~\cite{kitagawa_exploring_2010,kitagawa2012observation,kitagawa2012topological,PhysRevB.88.121406,PhysRevB.86.195414,PhysRevA.89.042327}. 

A connection between quantum walks and topology has been made by mapping the unitary evolution of the DTQW protocol to stroboscopic evolution under an effective Hamiltonian.
In certain cases---distinguished by a combination of symmetry and dimensionality---the effective Hamiltonian's bandstructure  exhibits a quantized invariant, analogous to those found in topological insulators \cite{kitagawa_exploring_2010,PhysRevB.86.195414,kitagawa2012observation,kitagawa2012topological}. 
On one hand, these invariants have helped to enable a sharp classification of non-interacting topological phases, which unlike conventional symmetry breaking phases, do not exhibit any local order parameters \cite{kitaev2009periodic,fidkowski2011topological}. 
On the other, they underly a multitude of exotic physical phenomena ranging from protected edge modes and quantized conductance to fractional charges and magnetic monopoles \cite{hasan_textitcolloquium_2010,PhysRevB.23.5632}.
Despite their importance and owing to their \emph{non-locality}, bulk topological invariants have been directly probed in only a handful of quantum optical systems \cite{atala2013direct, aidelsburger2015measuring,Mittal2016,PhysRevLett.115.040402} and a generic blueprint for their measurement remains an outstanding challenge. 

\begin{figure}
	\centering
		\includegraphics[width=3.4in]{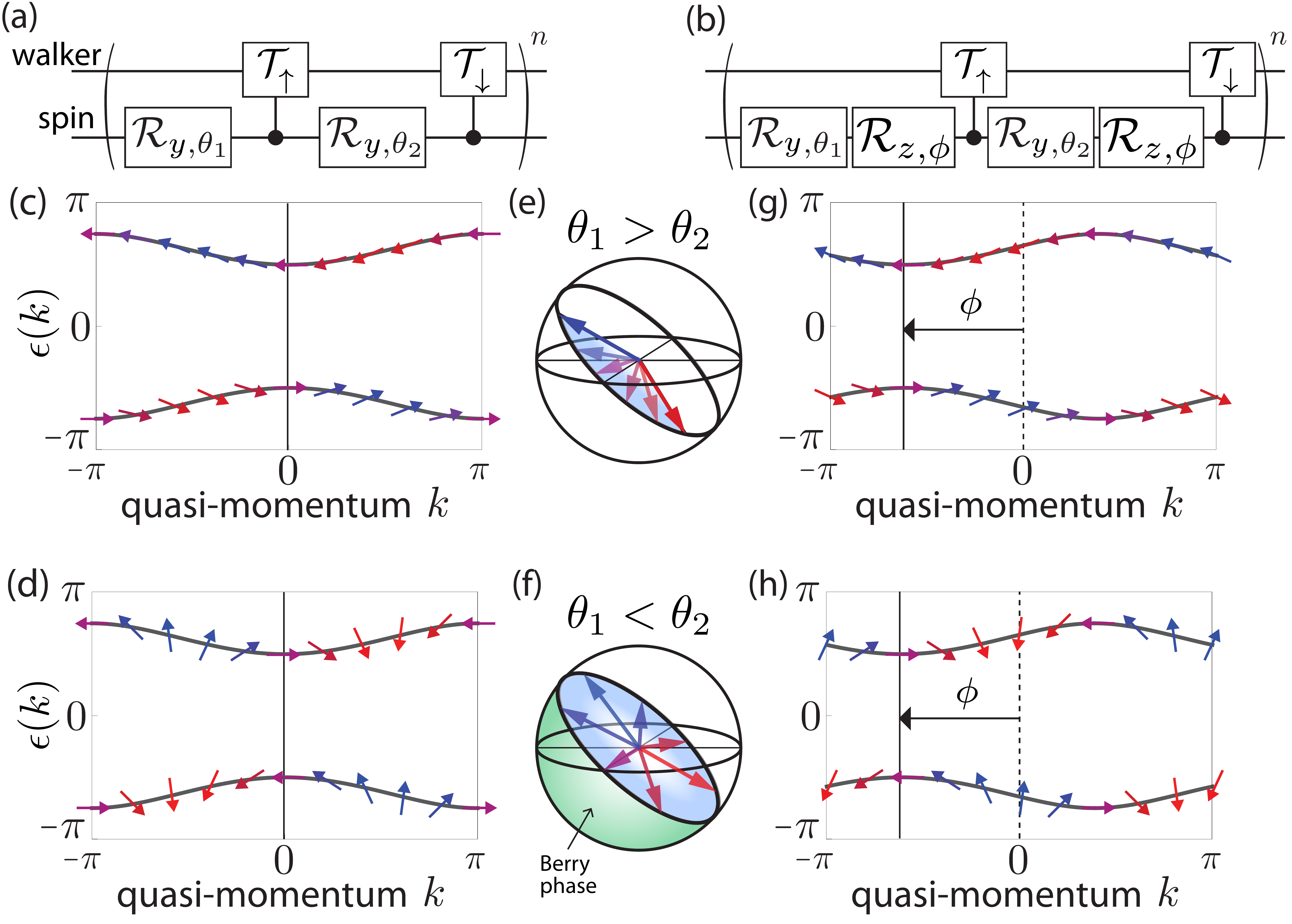}
	\caption{The sequence of unitary operations associated with a single step of  (a) the split-step quantum walk and (b) the Bloch-oscillating quantum walk. (c) The bandstructure and spin texture (arrows accompanying the band) for a trivial split-step quantum walk ($U_{SS}(3\pi/4,\pi4)$) with  $W=0$ and (d) for a topological split-step quantum walk ($U_{SS}(\pi/4,3\pi4)$) with $W=1$. (e,f) Schematic evolution of the spin eigenvectors in (c,d) as one traverses the Brillouin zone. In the topological phase, the spin texture fully winds around the origin as k varies from $[-\pi , \pi]$. (g,h) Analogous bandstructures for the trivial and topological Bloch-oscillating quantum walk. The shift in the effective momentum induced by the $z$-rotations (by $\phi$) are shown explicitly. 
	}
	\label{fig:fig1schem}
\end{figure}

In this Letter, we demonstrate that the simulation platform associated with discrete-time quantum walks is naturally suited for the direct extraction of topological invariants. 
We analyze a time-dependent,``Bloch-oscillating'' generalization of two classes (split-step and single-step) of one-dimensional DTQWs. In these protocols, a geometric signature of the topological invariant is imprinted as a Berry phase on the quantum state of the particle. 
We demonstrate that this phase can be extracted and disentangled from other contributions via a simple interferometric protocol \cite{abanin2013interferometric,grusdt2014measuring}. 
Our results directly connect to previous seminal observations of the refocusing behavior of time-dependent quantum walks~\cite{banuls_quantum_2006,cedzich_propagation_2013,PhysRevA.93.032329} and provide a  physical explanation for such behavior in terms of dynamical and geometric phases. 
While our approach is general, we propose an experimental realization in a circuit quantum electrodynamics architecture, leveraging the use of cavity-Schrodinger cat states to directly measure the topological invariant via Wigner tomography. 

\emph{General approach}---Let us begin by considering the family of protocols, dubbed split-step quantum walks, which act on a single spin-1/2 particle ($\{\ket{\downarrow},\ket{\uparrow}\}$) on a one-dimensional lattice $\{\ket{x}, x\in \mathbb{Z}\}$~\cite{kitagawa_exploring_2010}.  Parameterized by angles $\theta_1$ and $\theta_2$, these protocols consist of a sequence of unitary operations (Fig.~1a):~1) a spin rotation $R_y(\theta_1) = e^{-i\theta_1\sigma_y/2}$, 2) a spin-dependent translation $T_\uparrow= \sum_x \left[\ket{x+1}\bra{x} \otimes \ket{\uparrow}\bra{\uparrow} + \ket{x}\bra{x} \otimes \ket{\downarrow}\bra{\downarrow} \right]$, 3) a second spin rotation $R_y(\theta_2)$, and 4) a second spin-dependent translation  $T_\downarrow =  \sum_x \left[\ket{x}\bra{x} \otimes \ket{\uparrow}\bra{\uparrow} + \ket{x-1}\bra{x} \otimes \ket{\downarrow}\bra{\downarrow} \right]$ \cite{kitagawa_exploring_2010,kitagawa2012topological,suppinfo}.  Denoted $U_{\mathrm{SS}}(\theta_1,\theta_2)$, this sequence comprises a single step of the quantum walk.   

Although the protocol is defined in discrete unitary steps, the subsequent evolution can be related to that produced by an effective Hamiltonian (at stroboscopic times), where $e^{-i\Heff }=U_{\mathrm{SS}}(\theta_1,\theta_2)$. In the quasi-momentum basis $\ket{k}=\frac{1}{\sqrt{2\pi}}\sum_x e^{-ikx}\ket{x}$, the spin-dependent translation operators are diagonal ($T _\uparrow= e^{i \hat{k}(\sigma_z-1)/2} $ and $T _\downarrow= e^{i \hat{k}(\sigma_z+1)/2}$), and thus, 
\begin{equation}
\Heff  = \epsop \nop\cdot\boldsymbol{\sigma},  
\label{eq:Heff1}
\end{equation}
where $\hat{k} = \int dk \ket{k}\bra{k} k$; $\eps$ characterizes the bandstructure; and $\n$ specifies the corresponding spinor eigenstate  (Fig. 1c,d).  
An underlying chiral symmetry of $U_{\mathrm{SS}}$ constrains   ${\bf n}(\hat{k})$ to lie on a great circle of the Bloch sphere~\footnote{Note that there exists a chiral symmetric time frame if one moves half of the first rotation to the end of the walk cycle \cite{suppinfo}}. The number of times, $W$, which ${\bf n}(\hat{k})$ winds around the origin as $k$ varies from $[-\pi , \pi]$  defines the topological invariant of the walk~\cite{kitagawa_exploring_2010}.  
Depending on $\{ \theta_1$, $\theta_2\}$, the split-step quantum walk mimics motion either in a trivial band with winding number zero or a topological band with winding number unity (Fig.~1e,f). 
A key signature of this topological invariant is the geometric Berry phase, $\phi_\mathrm{geo} = \pi W$, acquired by the particle's wavefunction upon an adiabatic traversal  through the Brillouin zone. 

In order to imprint this Berry phase on the wavefunction of the quantum walker, we consider a time-dependent modification to $U_{\textrm{SS}}$, aimed at generating dynamics analogous to solid-state Bloch oscillations~\cite{banuls_quantum_2006,cedzich_propagation_2013}; the modified $m^{th}$ step unitary (Fig.~1b) is
\begin{equation}
U_{\textrm{SS}}^{(m)}(\theta_1,\theta_2) = T_\downarrow R_z(-m\phi) R_y\left(\theta_2\right)
T_\uparrow R_z(-m\phi) R_y\left(\theta_1\right) ,
\end{equation}
where $R_z(-m\phi)=e^{i\sigma_z m\phi/2}$ and $\phi = 2\pi/N$ for $N \in \mathbb{Z}$. 
Since $T_\uparrow R_z(-m\phi) = e^{im\phi/2}e^{i (\hat{k}+m\phi)(\sigma_z-1)/2}$ and $T_\downarrow R_z(-m\phi) = e^{-im\phi/2}e^{i (\hat{k}+m\phi)(\sigma_z+1)/2}$, the additional $z$-rotations simply shift the original quasi-momentum by a step-dependent amount and result in a modified effective Hamiltonian, 
\begin{equation}
\hat{H}_{\mathrm{eff}}^{(m)} = \epsilon_{\theta_1,\theta_2}(\hat{k}+m\phi) \textbf{n}_{\theta_1,\theta_2}(\hat{k}+m\phi)\cdot\boldsymbol{\sigma}.
\label{eq:HeffTD}
\end{equation}
In the   limit $\phi \ll 1$, this shift defines an adiabatic translation of momentum space, where the quantum walker traverses the full Brillouin zone in precisely $N$ steps.

\begin{figure}
	\centering
		\includegraphics[width=3.4in]{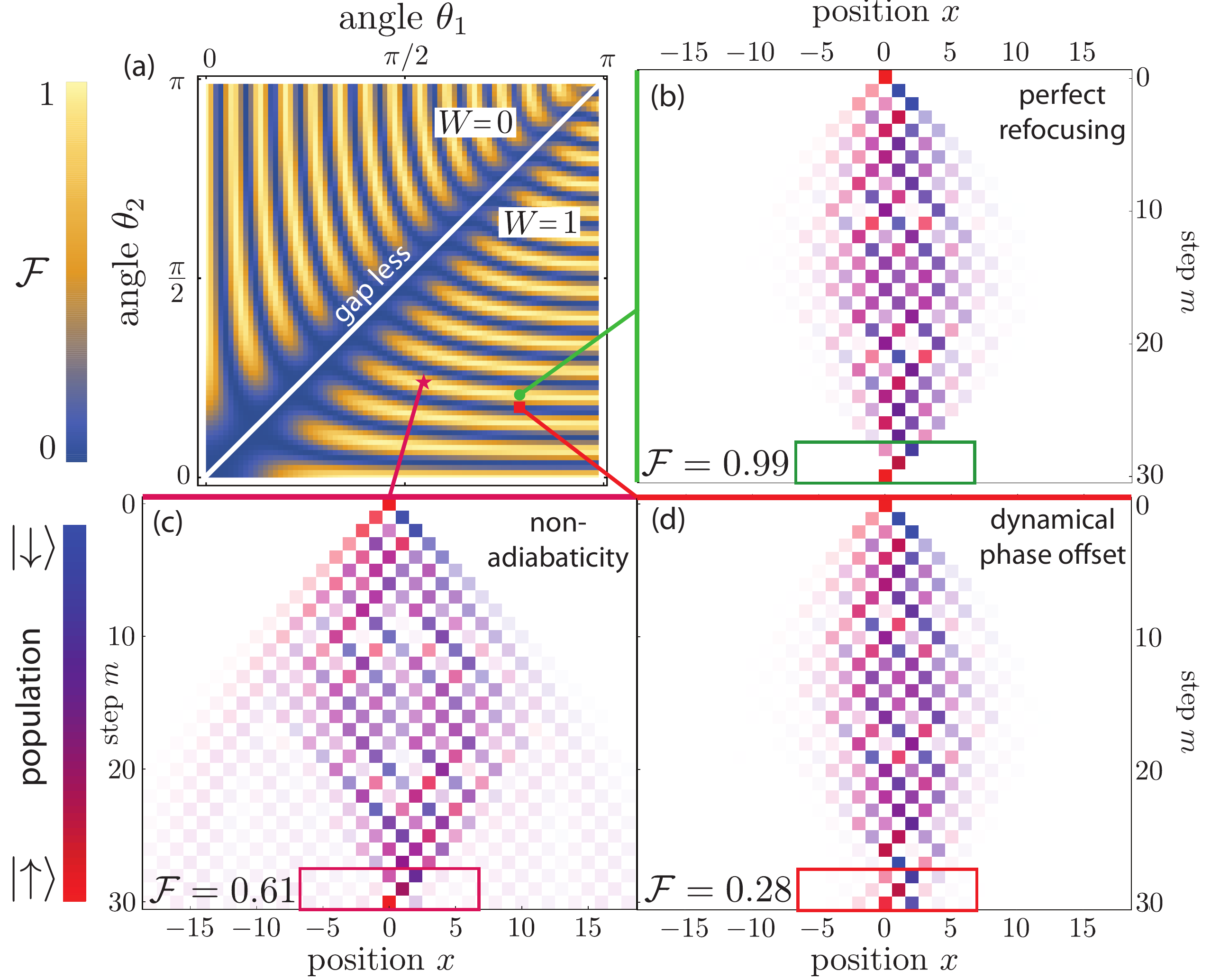}
	\caption{%
	(a) Refocusing fidelity $\mathcal{F} = |\braket{\psi_0|\psi_f}|^2$  computed from numerical simulations of an $N=30$ step Bloch-oscillating quantum walk as a function of  $\{ \theta_1$, $\theta_2 \}$, where $\phi = 2\pi / N$. 
	Two topologically distinct regions $W=0$ ($\theta_1>\theta_2$) and  $W=1$ ($\theta_1<\theta_2$) are separated by a gapless line. In the vicinity of this gap closure, the refocusing fidelity drops dramatically owing to non-adiabatic transitions. The observed stripe pattern in the refocusing fidelity follows the contour lines for which the accumulated dynamical phase is a multiple of $\pi$.
	(b)-(d)  Three specific time evolutions associated with various $\{ \theta_1$, $\theta_2 \}$: (b)  a perfectly refocusing walk, (c) a non-refocusing walk due to non-adiabatic transitions, and (d) a non-refocusing walk due to an accumulated dynamical phase of $\pi/2$.
	\label{fig:fig1_v1_schem}}
\end{figure}

To understand the dynamics of the ``Bloch-oscillating'' quantum walk, we map the discrete evolution associated with the series of step-dependent unitaries, $U_{\textrm{SS}}^{(m)}$, to continuous evolution under a time-dependent Schr\"odinger equation: $i\partial_t|\psi\rangle = H_{\textrm{eff}}(\hat{k}+\Delta k(t))|\psi\rangle$, where  $H_{\textrm{eff}}(\hat{k}+\Delta k(t))$ captures the step-dependent effective Hamiltonian in Eq.~(3) via  $\Delta k(t) = \phi\sum_m \Theta(t-m)$, where $\Theta$ is the Heaviside step function.

The analogy to Bloch oscillations is best captured by moving into a non-uniformly accelerating frame via the transformation, 
$U_{\Delta k(t)}=e^{i\hat{x}\Delta k(t)}$,
wherein the state, $\ket{\tilde{\psi}} = U_{\Delta k(t)}^\dagger \ket{\psi}$ satisfies
\begin{equation}
i\partial_t \ket{\tilde{\psi}}=\left(H_{\textrm{eff}}(\hat{k})+\hat{x} \phi \sum_m \delta(t-m)\right)\ket{\tilde{\psi}}.
\label{eq:TDSchro}
\end{equation}
The above time-evolution mirrors that of a particle on a stationary lattice receiving periodic kicks of magnitude $\phi$ and the resulting dynamics resemble Bloch oscillations. 

To see this, let us consider an initial state $\ket{\tilde{\psi} (0)} = \ket{k} \otimes \ket{\textbf{n}_{k}^\pm}$, where $\ket{\textbf{n}_{k}^+}$ and $\ket{\textbf{n}_{k}^-}$ are the spinor eigenstates (at momentum $k$) of  the upper and lower bands (Fig.~1), respectively. For $\phi \ll 1$,  the adiabatic theorem allows one to explicitly solve Eq.~(4) \cite{suppinfo},
\begin{equation}
\ket{\tilde{\psi} (t)} =   e^{i\phi_{\mathrm{dyn},\pm}}e^{i\phi_{\mathrm{geo},\pm} } \ket{k+m\phi}\ket{\textbf{n}_{k+m\phi}^\pm}.
\end{equation}
The momentum and spinor eigenstates simply follow their adiabats while the overall wavefunction acquires both a dynamical  and geometric phase,
\begin{eqnarray}
\phi_{\mathrm{dyn},\pm} &=& \pm \sum_{m\ge 0}\epsilon(k+m\phi) \nonumber \\ 
\phi_{\mathrm{geo},\pm} &=& i\phi\sum_{m\ge 0}\braket{\textbf{n}_{k+m\phi}^\pm|\partial_k \textbf{n}_{k+m\phi}^\pm}.
\label{eq:DiscretePhases}
\end{eqnarray}
Since $\ket{k+2\pi}\ket{\textbf{n}_{k+2\pi}^\pm} = \ket{k}\ket{\textbf{n}_{k}^\pm}$,  $\ket{\tilde{\psi} (m) }$ exhibits a recurrence to its initial state---up to a global phase---whenever $m/N$ is an integer \cite{banuls_quantum_2006,cedzich_propagation_2013}; this is precisely analogous to Bloch oscillations, in which Bloch waves recover their initial momentum upon any full traversal of the Brillouin zone.   

This recurrence forms the basis of our protocol to  measure topological invariants in quantum walks.  By performing an interference measurement (e.g. Ramsey spectroscopy) between the refocused wavefunction, $\ket{\tilde{\psi} (m)}$, and a reference state, one can directly extract the overall global phase $\phi_T = \phi_\mathrm{dyn} + \phi_\mathrm{geo}$. As will be shown below, it is possible to disentangle the dynamical and geometric contributions to $\phi_T$ by simply varying the overall step number.  In this way, one can extract $\phi_\mathrm{geo}$, thereby directly measuring the topological winding number.

\begin{figure}
	\centering
		\includegraphics[width=2.5in]{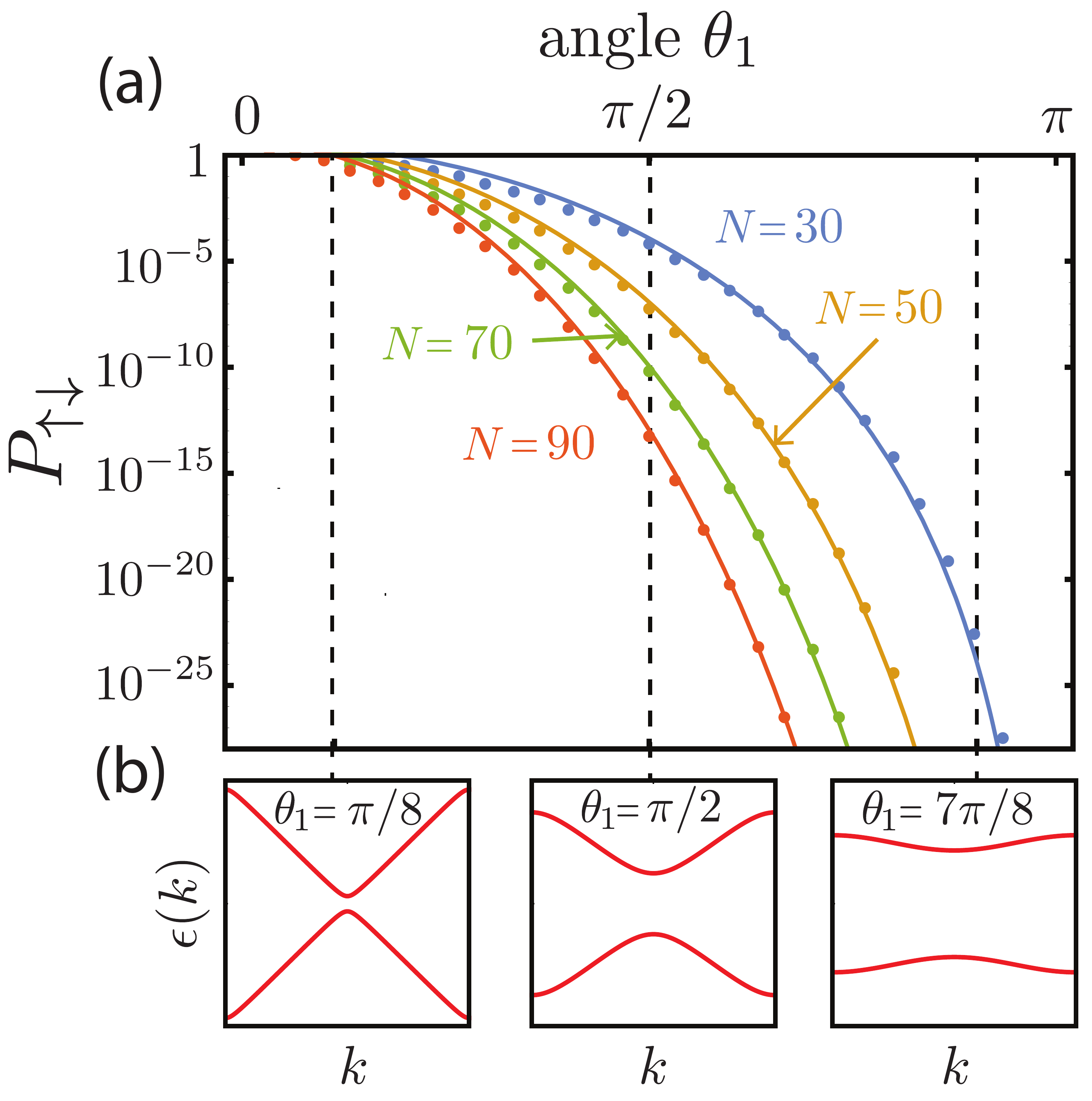}
	\caption{%
	(a) Non-adiabatic transition probability as a function of $\theta_1$ for the single-step quantum walk ($\theta_2=0$). Solid lines correspond to analytic formulae derived in \cite{cedzich_propagation_2013}, which capture the deviations from ideal refocusing behavior. Points correspond to $P_{\uparrow\downarrow}$ as computed from Eq.~(9), demonstrating that the physical origin of such deviations is non-adiabatic Landau-Zener transitions. 
	(b) Schematic bandstructures for various $\theta_1$. The bandgap increases as $\theta_1$ varies from $0$ to $\pi$ leading to a smaller non-adiabatic transition probability.   }
	\label{fig:Supp1}
\end{figure}

Although recurrence always occurs for initial momentum/spinor eigenstates (e.g.~$\ket{\tilde{\psi} (0)} = \ket{k} \otimes \ket{\textbf{n}_{k}^\pm}$), quantum walks are typically initialized with the particle localized at a single initial site.  As such states  consist of superpositions of eigenstates in both the upper and lower energy bands,
\begin{equation}
\ket{\tilde{\psi} (0)} =  \sum_k c_k \ket{k}\ket{\textbf{n}_{k}^+} + d_k \ket{k}\ket{\textbf{n}_{k}^-},
\label{eq:Decomp}
\end{equation} 
their refocusing behavior is significantly more subtle, requiring not only that each constituent eigenstate return to itself, but also that the total accrued phase is identical for all components.  
While the geometric phase acquired after $N$ steps is $\pi W$  for all eigenstates, the dynamical phase acquired by states in the upper and lower bands are opposite [Eq.~\ref{eq:DiscretePhases}]. Thus, the final state  will generally \emph{not} refocus to the initial state (Fig.~2a) and the wavefunction will remain spread over a number of sites (Fig.~2d). 

Fortunately, one can always ensure near-perfect refocusing (i.e.~enforcing a dynamical phase which is arbitrarily close to a multiple of $2\pi$ \footnote{We note that refocusing also occurs for dynamical phases that are an odd multiple of $\pi$. In such cases, one simply needs to account for the dynamical phase when calculating the winding number $W$ or double the total number of steps.}) by first characterizing the  fidelity as a function of  total step number.  
In particular, in the limit of large step number, the dynamical phase becomes proportional to $N$: $\phi_{\mathrm{dyn}} \approx N \times \bar{\epsilon}$, 
where $ \bar{\epsilon} = \int dk~ \epsilon (k )/2\pi$ \footnote{This is analogous to continuous-time evolution, where  $\phi_{\mathrm{dyn}}$ is directly proportional to the time taken to traverse a path in parameter space. }.  The refocusing fidelity, $\mathcal{F} = |\braket{\psi_0|\psi_f}|^2$,  is then given by \cite{suppinfo}
\begin{equation}
\mathcal{F} = \cos^2(N{\bar{\epsilon} }),
\end{equation}
enabling one to control the refocusing via a choice of $N$;  this is illustrated by the perfectly refocused state in Fig.~2b, where the global phase contains only the geometric component.

In addition to dynamical phase accumulation, non-adiabatic transitions can also lead to a lack of refocusing. This is particularly evident near regions where the gap closes as one transitions from a topological to trivial bandstructure. In such cases, even when $\phi_{\mathrm{dyn}} \propto 2\pi$~\footnote{While a physical interpretation of the dynamical phase is not obvious in the non-adiabatic case, it can be defined formally by solving the Schrodinger equation neglecting the band-mixing term, as is done in the supplement.}, the refocusing fidelity can be imperfect owing to interband Landau-Zener transitions (Fig.~2c). 

To quantify this effect, we consider single-step Bloch-oscillating quantum walks ($\theta_2=0$ in Eq.~(2)).  
As shown in Fig.~3, the effective bandstructure of the quantum walk changes as one varies  $\theta_1$, with the bandgap increasing continuously as $\theta_1$ varies from $0$ to $\pi$. An enhanced gap should decrease the non-adiabatic transition probability, $P_{\uparrow\downarrow}$, which can be explicitly computed for $\phi \ll 1$ as  ~\cite{suppinfo} 
\begin{equation}
P_{\uparrow\downarrow}\approx \phi^2\left|\sum_{m=1}^N \braket{\textbf{n}_{k+m\phi}^-|\partial_k \textbf{n}_{k+m\phi}^+}e^{-2 i \sum_{p=0}^m\epsilon(k+p\phi)}\right|^2.
\end{equation}
One finds that $P_{\uparrow\downarrow}$ is in quantitative agreement with analytics on single-step Bloch-oscillating quantum walks (Fig.~3) \cite{banuls_quantum_2006,cedzich_propagation_2013,suppinfo}.  As in the case of dynamical phase accumulation, one can tune the number of steps $N$ to minimize non-adiabatic refocusing errors below any desired threshold. 
 
So far we have shown how to construct a Bloch-oscillating quantum walk from an arbitrary split- or single-step quantum walk.  By choosing the number of steps $N$ such that the state of the particle is refocused, one finds that the final wavefunction differs from the initial state by only an imprinted geometric phase, which encodes the topology of the quantum walk.  While global phases are generally non-measurable, below we show how this geometric phase can be extracted interferometrically in a system with an additional internal state. 
 
\emph{Experimental realization}---We now propose an experimental blueprint for extracting topological invariants from Bloch-oscillating quantum walks in  a circuit quantum electrodynamics (cQED) architecture~\cite{PhysRevA.69.062320}. We consider a superconducting transmon qubit~\cite{koch_charge-insensitive_2007} coupled to a high-quality-factor electromagnetic cavity (Fig.~4a) and envision the quantum walk to take place in the phase space of the cavity mode  \cite{travaglione2002implementing}.  Each lattice site corresponds to a particular coherent state of the cavity  and the two logical states of the transmon ($ |g\rangle,|e\rangle $) form the internal spin of the walker~\cite{PhysRevA.78.042334}.

Spin rotations $R_y(\theta)$ and $R_z(\phi)$ can be performed using coherent microwave driving, with state-of-the-art pulse shaping techniques enabling single-qubit X and Y Clifford gates with greater than 99.9\% fidelity in as little as 20 ns~\cite{PhysRevLett.116.020501}.  Spin-dependent translations arise naturally from  the dispersive coupling  between the qubit  and the cavity,
$H_{\mathrm{int}} = \hbar(\chi/2) a^\dagger a\sigma_z$ (Fig.~4b)~\cite{Schuster2007}. Here,  $\sigma_z$ is the Pauli $z$-operator of the transmon qubit, $a^\dagger$ ($a$) the cavity raising (lowering) operator, and $\chi$ the strength of the qubit-cavity dispersive coupling.  
\begin{figure}
	\centering
		\includegraphics[width=3.4in]{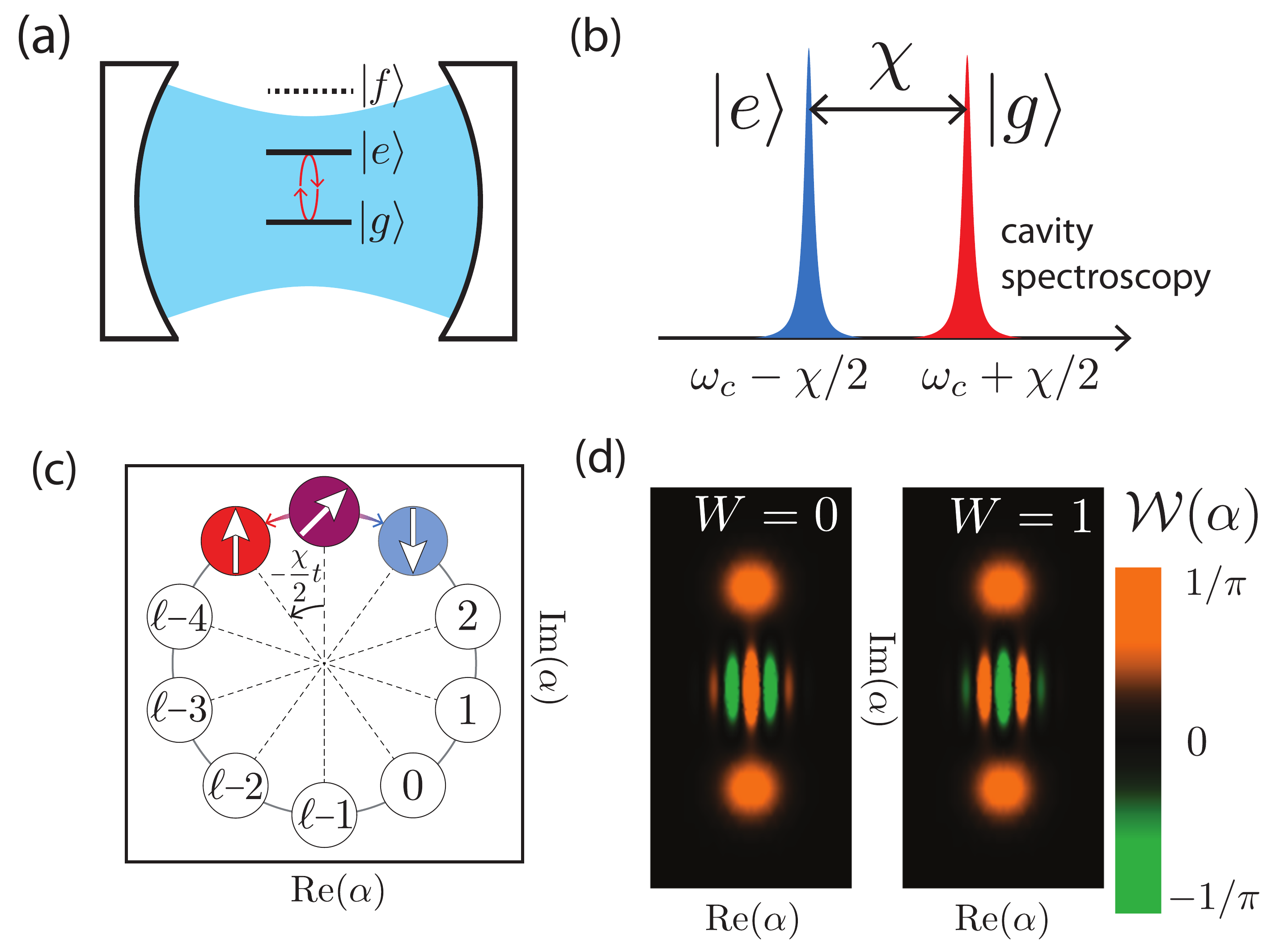}
	\caption{%
    (a) Schematic of the proposed cQED setup for realizing Bloch-oscillating quantum walks, utilizing a superconducting cavity mode coupled to a transmon qubit. The levels $ |g\rangle,|e\rangle $ form the internal spin states of the walker, while $|f\rangle $ is used as a shelving state.  (b) The qubit and cavity couple dispersively, realizing a  state-dependent shift of the bare cavity transition frequency, $\omega_c$, that naturally enables spin-dependent translations.  (c) The quantum walk takes place on a circular lattice in the phase space of the cavity. Each coherent state depicted in the figure represents a particular lattice site of the walk.  Spin down (up) corresponds to state $ |g\rangle$ ($|e\rangle $). (d) Wigner tomography, $\mathcal{W}(\alpha)$, of the cavity following a refocused Bloch-oscillating quantum walk in the topological and trivial bandstructures reveals the underlying winding number in the phase of the interference fringes.  
}
	\label{fig:fig1_v1_schem}
\end{figure}

In combination, the above operations enable the realization of a quantum walk on a circular lattice in cavity phase space (Fig.~4c). In particular, one 
 initializes the cavity in a coherent state $|\alpha\rangle$, with the qubit in the ground state $|g\rangle$.
 After applying the desired unitary rotation to the qubit, a waiting period of time $t$ allows the dispersive interaction to naturally implement the spin-dependent translation.  Indeed, a coherent state $|\alpha\rangle$ in the cavity frame precesses either clockwise ($|\alpha\rangle |e\rangle \rightarrow|\alpha \exp{(i\chi t/2 )}\rangle  |e\rangle $) or counterclockwise ($|\alpha\rangle |g\rangle \rightarrow|\alpha \exp{(-i\chi t/2 )}\rangle  |g\rangle $) depending on the qubit state (Fig.~4b,c).  Choosing $t$ such that $\chi t = 2\pi / L$ defines the  $L$ coherent state ``lattice sites'': $\{|\alpha \exp{(i2\pi \ell /L)}\rangle, \ell \in[0,L-1]\}$. 

These two basic steps (unitary rotation and spin-dependent translation) can then be repeated  to realize a Bloch-oscillating quantum walk.   Measurement of the quantum walker's spin and position after each step can be performed via full tomography of the cavity-qubit system~\cite{vlastakis_deterministically_2013}.  

To directly probe the topological invariant via the imprinted geometric phase, one must perform interferometry between the refocused wavefunction and a reference state. 
This is naturally enabled by the proposed cQED architecture, where one can initialize the system  in a cavity Schr\"odinger cat state, corresponding to a coherent superposition,  $1/\sqrt{2}(|\alpha\rangle |g\rangle + |0\rangle |f\rangle)$, where $|0\rangle$ is the vacuum state of the cavity and  $|f\rangle$ is the second excited state of the qubit. Crucially, the $|0\rangle |f\rangle$ state behaves as a  phase reference since it is immune to  both the unitary spin rotations and the dispersive coupling. The $\ket{f}$ state in transmon qubits can exhibit coherence and decay times in excess of $20~\mu s$~\cite{PhysRevLett.114.010501}, while the aforementioned pulse-shaping techniques result in off-resonant leakage errors  $< 10^{-5}$.  

Upon refocusing of the $|\alpha\rangle |g\rangle $ component, the final wavefunction takes the form: $1/\sqrt{2}(e^{i \pi W}|\alpha\rangle |g\rangle + |0\rangle |f\rangle)$ and the topological winding number manifests in the geometric relative phase between the two components.  After disentangling the spin and cavity degrees of freedom via number-selective qubit pulses (i.e.~$|0\rangle |f\rangle \rightarrow |0\rangle |g\rangle$)~\footnote{By using a single local oscillator to perform both excitation into and out of the $\ket{f}$ state, the evolved dynamical phase associated with this state is automatically kept track of and cancelled.}, one can perform full Wigner tomography of the cavity state. As illustrated in Fig.~4d, the resulting interference patterns display fringes whose phase corresponds to $ \pi W$ \cite{vlastakis_deterministically_2013, ramasesh2016}. 

In summary, we have demonstrated that the simulation platform associated with quantum walks can enable the direct measurement of bulk topological invariants. In particular, by constructing Bloch-oscillating analogues of both split- and single-step quantum walks, we have introduced an interferometric protocol to directly measure the winding number associated with a quantum walk's effective bandstructure. A key feature of such Bloch-oscillating quantum walks is their natural refocusing behavior, whose microscopic origin arises from an interplay between dynamical and geometric phases as well as non-adiabatic transitions.  Looking forward, our results can be directly extended to measurements of  quantum walk topological invariants in higher dimensions, and provide a bridge toward probing many-body invariants  associated with interacting quantum walks.  

We gratefully acknowledge the insights of and discussions with E. Demler, L. Martin, S. Hacohen-Gourgy, and C. Navarrete-Benlloch. This research is supported in part by the Miller Institute for Basic Research in Science, the U.S. Army Research Office (ARO) under grant no. W911NF-15-1-0496 and by the AFOS under grant no. FA9550-12-1-0378.    M.R. gratefully acknowledges the Villum Foundation and the People Programme (Marie Curie Actions) of the European Unions Seventh Framework Programme (FP7/2007-2013) under REA grant agreement PIIF-GA- 2013-627838 for support.  VVR acknowledges funding via a NSF graduate student fellowship.  
\bibliography{My_Library}

\pagebreak
\clearpage
\widetext
\begin{center}
\textbf{\large Supplemental Material for Direct Probe of Topological Invariants Using Bloch Oscillating Quantum Walks}

{\large V. V. Ramasesh, E. Flurin, M. Rudner, I. Siddiqi, N. Y. Yao}
\end{center}

\setcounter{equation}{0}
\setcounter{figure}{0}
\setcounter{table}{0}
\setcounter{page}{1}
\makeatletter
\renewcommand{\theequation}{S\arabic{equation}}
\renewcommand{\thefigure}{S\arabic{figure}}

\section{Adiabatic limit of Bloch-oscillating Quantum Walk}
Here, we provide theoretical details illustrating the adiabatic limit of the Bloch-oscillating quantum walk protocol, even in the presence of discrete jumps in the effective Hamiltonian. In the accelerating frame, the Schr\"odinger equation transforms into
$i \partial_t \ket{\tilde{\psi}}=(\tilde{H}-i U_{\Delta k(t)}^\dagger\partial_t U_{\Delta k(t)})\ket{\tilde{\psi}}$, where
$\tilde{H}=U_{\Delta k(t)}^\dagger H(t) U_{\Delta k(t)}=H_{\mathrm{eff}}(\hat{k})$.
Inserting the time-dependence of $\Delta k(t)$ yields:
\begin{equation}
i\partial_t \ket{\tilde{\psi}}=\left(H_\mathrm{eff}(\hat{k})+\hat{x} (\partial_t \Delta k)\right)\ket{\tilde{\psi}}.
\label{eq:TDSchro}
\end{equation}
In the non-accelerating lab frame, an initial plane-wave state $\ket{k}$ is stationary.  In the accelerating frame,  plane wave states evolve through the Brillouin zone: $\ket{k}\rightarrow\ket{k-\Delta k}$.     Thus, one can consider a (completely general) ansatz for $\ket{\tilde{\psi}}$:
\begin{equation}
\begin{split}
\ket{\tilde{\psi}}=\frac{1}{2\pi}\int_{\mathrm{BZ}}dk~\ket{k+\Delta k(t)}\left(c_k(t)\ket{\textbf{n}_{k+\Delta k(t)}^+}+d_k(t)\ket{\textbf{n}_{k+\Delta k(t)}^-}\right).
\end{split}
\end{equation}
To solve for $c_k(t)$ and $d_k(t)$, we start by evaluating the LHS of eq.~(\ref{eq:TDSchro}) with the above ansatz:
\begin{eqnarray*}
i\partial_t\ket{\tilde{\psi}} = \frac{i}{2\pi}\int_{\mathrm{BZ}}dk~\partial_t \left[ \ket{k+\Delta k(t)}\left(c_k(t)\ket{\textbf{n}_{k+\Delta k(t)}^+}+d_k(t)\ket{\textbf{n}_{k+\Delta k(t)}^-}\right)\right] \\
= \frac{i}{2\pi}\int_{\mathrm{BZ}}dk~\ket{k+\Delta k(t)}\left[ (\partial_t{c}_k(t))\ket{\textbf{n}_{k+\Delta k(t)}^+}
+(\partial_t{d}_k(t))\ket{\textbf{n}_{k+\Delta k(t)}^-} \right]\\
+ \frac{i}{2\pi}\int_{\mathrm{BZ}}dk~\ket{k+\Delta k(t)}\left[ {c}_k(t)\partial_t\ket{\textbf{n}_{k+\Delta k(t)}^+}
+{d}_k(t)\partial_t\ket{\textbf{n}_{k+\Delta k(t)}^-} \right]\\
+ \frac{i}{2\pi}\int_{\mathrm{BZ}}dk~(\partial_t \ket{k+\Delta k(t)})\left[ {c}_k(t)\ket{\textbf{n}_{k+\Delta k(t)}^+}
+{d}_k(t)\ket{\textbf{n}_{k+\Delta k(t)}^-} \right].
\end{eqnarray*}
The final term in the above expression can be rewritten as:
\begin{eqnarray}
\partial_t \ket{k+\Delta k(t)} &=& (\partial_t \Delta k) \partial_k \ket{k+\Delta k(t)} = (\partial_t \Delta k) \partial_k \left( \frac{1}{\sqrt{2\pi}} \sum_x e^{-ikx}\ket{x} \right) =(\partial_t \Delta k) \left(-i\hat{x} \ket{k+\Delta k(t)} \right).
\end{eqnarray}
We can thus simplify the expression for the LHS of eq.~(\ref{eq:TDSchro}):
\begin{eqnarray*}
i\partial_t\ket{\tilde{\psi}}
= \frac{i}{2\pi}\int_{\mathrm{BZ}}dk~\ket{k+\Delta k(t)}\left[ (\partial_t{c}_k(t))\ket{\textbf{n}_{k+\Delta k(t)}^+}
+(\partial_t{d}_k(t))\ket{\textbf{n}_{k+\Delta k(t)}^-} \right]\\
+ \frac{i}{2\pi}\int_{\mathrm{BZ}}dk~\ket{k+\Delta k(t)}\left[ {c}_k(t)\partial_t\ket{\textbf{n}_{k+\Delta k(t)}^+}
+{d}_k(t)\partial_t\ket{\textbf{n}_{k+\Delta k(t)}^-} \right]+\hat{x}(\partial_t \Delta k) \ket{\tilde{\psi}}
\end{eqnarray*}
Inspection reveals that the term $\hat{x}(\partial_t \Delta k)\ket{\tilde{\psi}}$ appears on both the LHS and RHS of the Schr\"odinger equation, so we can cancel this term, giving 
\begin{equation}
\begin{split}
\frac{i}{2\pi}\int_{\mathrm{BZ}}dk~\ket{k+\Delta k(t)}\left[ (\partial_t{c}_k(t))\ket{\textbf{n}_{k+\Delta k(t)}^+}
+(\partial_t{d}_k(t))\ket{\textbf{n}_{k+\Delta k(t)}^-} \right] \\
+ \frac{i}{2\pi}\int_{\mathrm{BZ}}dk~\ket{k+\Delta k(t)}\left[ {c}_k(t)\partial_t\ket{\textbf{n}_{k+\Delta k(t)}^+}
+{d}_k(t)\partial_t\ket{\textbf{n}_{k+\Delta k(t)}^-} \right] =  H_\mathrm{eff}(\hat{k})\ket{\tilde{\psi}}.
\end{split}
\end{equation}
At this point, all of the plane-wave states $\ket{k}$ with differing momenta are uncoupled. Projecting onto a particular plane wave state:
\begin{equation}
(i\partial_t{c}_k(t))\ket{\textbf{n}_{k+\Delta k(t)}^+}+(i\partial_t{d}_k(t))\ket{\textbf{n}_{k+\Delta k(t)}^-}=H_\mathrm{eff}(k)\ket{\tilde{\psi}} - i{c}_k(t)\partial_t\ket{\textbf{n}_{k+\Delta k(t)}^+}
-i{d}_k(t)\partial_t\ket{\textbf{n}_{k+\Delta k(t)}^-}.
\end{equation}
Further projecting  onto the states $\ket{\textbf{n}^\pm_{k+\Delta k(t)}}$ gives two coupled equations for $c_k$ and $d_k$:
\begin{eqnarray}
i \dot{c}_k(t) &=& \epsilon(k+\Delta k)c_k(t) - i(\partial_t \Delta k)\braket{\textbf{n}^+_{k+\Delta k(t)}|\partial_k| \textbf{n}^+_{k+\Delta k(t)}}c_k(t) -i(\partial_t \Delta k)\braket{\textbf{n}^+_{k+\Delta k(t)}|\partial_k| \textbf{n}^-_{k+\Delta k(t)}}d_k(t) \label{eq:C} \\
i \dot{d}_k(t) &=& -\epsilon(k+\Delta k)d_k(t) - i(\partial_t \Delta k)\braket{\textbf{n}^-_{k+\Delta k(t)}|\partial_k| \textbf{n}^-_{k+\Delta k(t)}}d_k(t) -i(\partial_t \Delta k)\braket{\textbf{n}^-_{k+\Delta k(t)}|\partial_k| \textbf{n}^+_{k+\Delta k(t)}}c_k(t).
\label{eq:D}
\end{eqnarray}
The three terms on the RHS of eqs.~(\ref{eq:C},\ref{eq:D}) correspond to, respectively, the dynamical phase, the Berry phase, and the non-adiabatic mixing between bands.  We neglect the mixing term in our initial analysis (see the next section for details), which  decouples the equations for $c_k(t)$ and $d_k(t)$, yielding  solutions
\begin{eqnarray}
c_k(t)=\exp\big[\int_0^t -i\epsilon(k+\Delta k(\tau))-(\partial_\tau \Delta k)\braket{\textbf{n}^+_{k+\Delta k(\tau)}|\partial_k|\textbf{n}^+_{k+\Delta k(\tau)}}  d\tau\big] c_k(0) \\
d_k(t)=\exp\big[\int_0^t i\epsilon(k+\Delta k(\tau))-(\partial_\tau \Delta k)\braket{\textbf{n}_{k+\Delta k(\tau)}^-|\partial_k |\textbf{n}^-_{k+\Delta k(\tau)}}  d\tau\big] d_k(0).
\end{eqnarray}
Performing the integration gives the dynamics as a function of the step number $m$:
\begin{eqnarray}
c_k(m)=\exp\big[&-i\sum_{m\ge 0} \epsilon(k+m\phi)-\phi \sum_{m>0}\braket{\textbf{n}^+_{k+m\phi}|\partial_k|\textbf{n}^+_{k+m\phi}}\big]c_k(0)\\
d_k(m)=\exp\big[&i\sum_{m\ge 0} \epsilon(k+m\phi)-\phi \sum_{m>0}\braket{\textbf{n}^{-}_{k+m\phi}|\partial_k| \textbf{n}_{k+m\phi}^-}\big]d_k(0).
\end{eqnarray}
The first sum in the exponential corresponds to the discretized dynamical phase $\phi_{\mathrm{dyn},\pm}^\Delta$, while the second corresponds to the discretized Berry phase $\phi_{\mathrm{Berry},\pm}^\Delta$:
\begin{eqnarray}
\phi_{\mathrm{dyn},\pm}^\Delta &=& \pm \sum_{m\ge 0}\epsilon(k+m\phi) \\
\phi_{\mathrm{Berry},\pm}^\Delta &=& i\phi\sum_{m\ge 0}\braket{\textbf{n}^\pm_{k+m\phi}|\partial_k| \textbf{n}^\pm_{k+m\phi}}.
\end{eqnarray}
When the number of steps in the walk is such that the entire Brillouin zone is traversed, and in the limit of small $\phi$ (or equivalently, large $N$), the discretized Berry phase approaches the continuous Berry phase $i\oint\braket{\textbf{n}^{\pm}_{k}|\partial_k \textbf{n}^\pm_{k}}dk$.  The dynamical phase can also be approximated in this limit as $\phi_{\mathrm{dyn},\pm}^\Delta\rightarrow\pm N/(2\pi)\oint\epsilon(k) ~dk$.

\subsection{Calculation of the non-adiabatic (Landau-Zener) transition probability}

Here, we derive an approximate formula for the interband transitions introduced by our time-dependent quantum walk protocol.  We again solve the Schr\"odinger equations for the $c_k(t)$ and $d_k(t)$ coefficients, this time keeping the final non-adiabatic terms.  In this calculation we estimate the contribution of interband mixing while staying close to the adiabatic limit.  To do this, we consider the situation (different from the main text) in which the state is initialized in a single $\ket{k}$ state and in the lower band so that $d_k$ is always close to unity and $c_k\ll 1$.  Instead of solving for $|c_k|^2$ directly, we solve for $c_k^*d_k$, since approximately $|c_k^* d_k|^2 \approx |c_k|^2$.  Equations \ref{eq:C} and \ref{eq:D} can be combined to read 
\begin{equation}
i\partial_t(c_k^* d_k) = -2\epsilon(k+\Delta k)c_k^* d_k - i\partial_t \Delta k \braket{\textbf{n}_{k+\Delta k}^-|\partial_k|\textbf{n}_{k+\Delta k}^+}(|c_k|^2 - |d_k|^2),
\end{equation}
which, by enforcing normalization ($|c_k|^2 + |d_k|^2 = 1$), transforms into 
\begin{equation}
i\partial_t (c_k^* d_k) = -2\epsilon(k+\Delta k)c_k^* d_k - i\partial_t \Delta k \braket{\textbf{n}_{k+\Delta k}^-|\partial_k|\textbf{n}_{k+\Delta k}^+}(2|c_k|^2-1).
\end{equation}
Assuming that the band mixing is small, so that $|c_k^* d_k|^2 \ll 1$, this equation becomes
\begin{equation}
\partial_t (c_k^* d_k e^{-2i\int_0^t d\tau ~ \epsilon(k+\Delta k)}) \approx \partial_t \Delta k \braket{\textbf{n}_{k+\Delta k}^- |\partial_k|\textbf{n}_{k+\Delta k}^+} e^{-2i\int_0^t d\tau \epsilon(k+\Delta k)},
\end{equation}
which can be directly integrated to yield 
\begin{equation}
|c_k^* d_k|(t) \approx \left| \int_0^t d\tau ~ (\partial_\tau \Delta k) \braket{\textbf{n}_{k+\Delta k}^-|\partial_k|\textbf{n}_{k+\Delta k}^+}  e^{-2i\int_0^\tau d\tau' \epsilon(k+\Delta k)}\right|.
\end{equation}
Inserting explicitly the time-dependence of $\Delta k (t)$---composed of step functions---gives $\delta$-functions in the integrand, transforming the integral into a discrete sum.  Here we specialize to a full traversal of the Brillouin zone (in $N$ steps), so 
\begin{equation}
|c_k^* d_k| \approx \phi \left| \sum_{m=1}^N \braket{\textbf{n}_{k+n\phi}^-|\partial_k|\textbf{n}^+_{k+m\phi}} e^{-2i\sum_0^m \epsilon(k+p\phi)} \right|.
\end{equation}
Thus, the approximate probability for interband transitions is given by
\begin{equation}
P_{\uparrow\downarrow}\approx \phi^2\left|\sum_{m=1}^N \braket{\textbf{n}^-_{k+m\phi}|\partial_k |\textbf{n}^+_{k+m\phi}}e^{-2 i \sum_0^m\epsilon(k+p\phi)}\right|^2.
\end{equation}
This probability is plotted in Figure 3 of the main text, showing clearly that even with a moderate number of steps our protocol is essentially adiabatic.

\section{Symmetries and implementation of the split-step quantum walk}

Here we clarify two points in our definition and implementation of the split-step quantum walk.  First, the existence of topological invariants in $U_{SS}(\theta_1,\theta_2)$ relies on the chiral symmetry of the walk operator.  To show that $U_{SS}(\theta_1,\theta_2)$ possesses a chiral symmetry as described in \cite{PhysRevB.88.121406}, we move into a time frame in which the walk unitary exhibits an inversion point.  Namely, one can move from the original time frame, in which $U_{SS}(\theta_1,\theta_2) = T_{\downarrow} R_y(\theta_2) T_{\uparrow} R_y(\theta_1)$, to the one in which the transformed operator $U_{SS}^{tf}(\theta_1,\theta_2)$ is $U_{SS}^{tf}(\theta_1,\theta_2) = R_y(\theta_1)^{1/2}T_{\downarrow} R_y(\theta_2) T_{\uparrow} R_y(\theta_1)^{1/2}$, exhibiting a clear inversion point.

Second, we clarify  the implementation of the walk operation $U_{SS}$ in the circuit QED setting.  As described in the main text, the spin-dependent translations are naturally implemented by the dispersive coupling between the cavity and qubit: $\hbar (\chi/2) a^\dagger a \sigma_z$.  In the rotating frame of the cavity, this evolution is given by
\begin{equation}
U_{\mathrm{disp}} = \sum_{l=0}^{L-1} \left(\ket{\alpha e^{i2\pi l/L }}\bra{\alpha e^{i 2\pi (l-1/2)/L}} \otimes \ket{g}\bra{g} + \ket{\alpha e^{i2\pi l/L }}\bra{\alpha e^{i 2\pi (l+1/2)/L}} \otimes\ket{e}\bra{e}\right).
\end{equation}
To see that this corresponds to the usual split-step quantum walk, let us re-write $U_\mathrm{disp}$ in two equivalent ways:
\begin{eqnarray}
U_{\mathrm{disp}} &=& \left[ \sum_l \left(\ket{\alpha e^{i2\pi l/L }}\bra{\alpha e^{i 2\pi (l-1)/L}} \otimes \ket{g}\bra{g}+\ket{\alpha e^{i2\pi l/L }}\bra{\alpha e^{i 2\pi l/L}} \otimes\ket{e}\bra{e} \right)\right] e^{i(\pi/L) a^\dagger a} \\
U_{\mathrm{disp}} &=& e^{-i(\pi/L) a^\dagger a}\left[\sum_l \left(\ket{\alpha e^{i2\pi l/L }}\bra{\alpha e^{i 2\pi l/L}} \otimes \ket{g}\bra{g}+\ket{\alpha e^{i2\pi l/L }}\bra{\alpha e^{i 2\pi (l+1)/L}} \otimes\ket{e}\bra{e} \right)\right].
\end{eqnarray}
Recognizing the operators in square brackets above as $T_\uparrow$ and $T_\downarrow$, respectively, one finds
\begin{equation}
U_{\mathrm{disp}} = T_\downarrow e^{i(\pi/L) a^\dagger a} =  e^{-i(\pi/L) a^\dagger a}T_\uparrow,
\end{equation}
immediately implying,
\begin{equation}
U_{\mathrm{disp}}R_y(\theta_2)U_{\mathrm{disp}}R_y(\theta_1) = T_\downarrow R_y(\theta_2) T_\uparrow R_y(\theta_1).
\end{equation}

\section{Relation between Bloch-oscillating quantum walks and previous refocusing proofs}

As described in the main text, our results directly connect to seminal previous observations that time-dependent single-step quantum walks can refocus under certain conditions.  Here, we demonstrate that the revival theorem proven in \cite{PhysRevA.93.032329}, can be recast  in terms of Bloch oscillations and the resulting dynamical and geometric phases.  

In the  notation of \cite{PhysRevA.93.032329}, the operator $W_{\Phi}^{[m,1]}$ stands for steps $1$ through $m$ of a Bloch-oscillating quantum walk which traverses the Brillouin zone in steps of $\Phi$; that is:
\begin{equation}
W_{\Phi}^{[m,1]} = W_\Phi (m)W_\Phi (m-1) . . . W_\Phi (1).
\end{equation}
In the notation of the main text,  $W_\Phi (m) = U_{SS}^{(m)}$.  The revival theorem  states that if $\Phi = 2\pi/m$, then 
\begin{eqnarray}
||W_\Phi^{[m,1]} - (-1)^{(m/2+1)}(\mathbb{I})||_{\mathrm{op}} &=& 2|a|^{m/2}, m~ \mathrm{even} \\
||W_\Phi^{[2m,1]} - (-\mathbb{I})||_{\mathrm{op}} &=& 2|a|^m, m~ \mathrm{odd} 
\end{eqnarray}
The parameter $a$ is related to the rotation angle: $a = \cos{(\theta/2)}$.  For our present purposes, all that matters is that $a \ll 1$, so that the RHS of both of the above equations tends to zero as $m\rightarrow\infty$,
\begin{eqnarray}
\lim_{m\rightarrow\infty} W_\Phi^{[m,1]} \approx (-1)^{(m/2+1)}\mathbb{I}, ~m ~\mathrm{even} \label{eq:ref1}\\
\lim_{m\rightarrow\infty} W_\Phi^{[2m,1]} \approx -\mathbb{I},~ m~\mathrm{odd} \label{eq:ref2}
\end{eqnarray}
Note that the operator $W_\Phi^{[m,1]}$ corresponds to a Bloch-oscillating walk in which a single traversal of the Brillouin zone is carried out with $m$ steps, while the operator $W_\Phi^{[2m,1]}$ corresponds to a walk which traverses the Brillouin zone twice, with each traversal comprising $m$ steps. 

\vspace{3mm}

\noindent Let us consider the accrued phases due to each of these protocols.  In the first case ($W_\Phi^{[m,1]}$), each state (regardless of band) acquires a geometric phase of $\pi$ due to the topology (recall that unlike the split-step quantum walk, all single-step quantum walks are in the topologically non-trivial phase with winding number $W=1$).  As for the dynamical phase, each traversal of the Brillouin zone gives a dynamical phase of $\pm m\pi/2$ (see below), corresponding to a global phase of either $\pi$ or $0$ depending on the parity of $m/2$.  So the total phase is $\pi(1+m/2)$, effecting a total multiplication by $(-1)^{(m/2+1)}$, as stated in eq.~(\ref{eq:ref1}).  For the odd case ($W_\Phi^{[2m,1]}$), the two traversals of the Brillouin zone give a geometric phase of $2\pi$.  However, the dynamical phase is $\pm m\pi$, which is always equivalent to $\pi$ since $m$ is odd.  In this case, the total imparted phase is always $\pi$, and thus the wavefunction is always multiplied by $-1$.

\vspace{3mm}

\noindent Let us now compute the dynamical phase, for a single Brillouin zone traversal, as a function of $N$ the total number of steps used (single-step case).  The quasienergy is given by $\epsilon(k) = \arccos{\left(\cos{(k)}\cos{(\theta/2)} \right)}$ ~\cite{kitagawa_exploring_2010} .
The dynamical phase $\phi_{\mathrm{dyn}}$ is calculated in the limit $N \gg 1$ as 
\begin{equation}
\phi_{\mathrm{dyn}} \approx \frac{N}{2\pi} \int_{-\pi}^{\pi}dk~\epsilon (k) = N\pi/2
\end{equation}

\section{Refocusing fidelity as a function of dynamical phase}
Here, we derive eq.~(8) of the main text, namely that the refocusing fidelity varies sinusoidally with the accrued phase.  
To see this, we write $\ket{\psi_0}$ in terms of $\ket{k}$ states:
$\ket{\psi_0} = \int dk~ (c_k\ket{k}\ket{\textbf{n}_k^+} + d_k\ket{k}\ket{\textbf{n}_k^-})$,
The state after the Bloch-oscillating quantum walk (apart from a global Berry phase which does not impact refocusing fidelity) is given by 
\begin{equation}
\ket{\psi_f} = \int dk~ (c_k e^{i\phi_{\mathrm{dyn}}}\ket{k}\ket{\textbf{n}_k^+} + d_k e^{-i\phi_{\mathrm{dyn}}} \ket{k}\ket{\textbf{n}_k^-}).
\end{equation}
The refocusing fidelity $|\braket{\psi_f|\psi_0}|^2$ is then 
\begin{equation}
|\braket{\psi_f|\psi_0}|^2 = \left| e^{i\phi_{\mathrm{dyn}}} \left( \int dk~ |c_k|^2 \right)+ e^{-i\phi_{\mathrm{dyn}}} \left( \int dk~|d_k|^2 \right) \right|^2 = \cos^2{(\phi_\mathrm{dyn})},
\label{eq:Overlap}
\end{equation}
where we have used a specific relation between the inner product of two spinor states and their corresponding Bloch-sphere vectors, namely that $|\langle \psi | \phi \rangle|^2 = (1/4)\left|\vec{n}+\vec{m}\right|^2$, where $\vec{n}$ and $\vec{m}$ point in the direction of the states $\phi$ and $\psi$.  Using this relation, one can easily show that (for the specific initial state localized on a single lattice site and in a $\sigma_z$ eigenstate) the two integrals in eq.~(\ref{eq:Overlap}) evaluate to $1/2$.

\end{document}